\documentclass{article}

\usepackage{arxiv}

\usepackage{cite}
\usepackage{amsmath,amssymb,amsfonts}
\usepackage[linesnumbered,ruled,vlined]{algorithm2e}
\usepackage{graphicx}
\usepackage{textcomp}
\usepackage{xcolor}
\usepackage{booktabs}
\usepackage{listings}
\usepackage{multirow}
\usepackage{framed}
\usepackage{makecell}
\usepackage{soul}
\usepackage[utf8]{inputenc}
\usepackage[T1]{fontenc}
\usepackage{microtype}
\usepackage{rotating}
\usepackage{paralist}
\usepackage{tabularx}
\usepackage{multicol}
\usepackage{pbox}
\usepackage{enumitem}	
\usepackage{colortbl}
\usepackage{pifont}
\usepackage{xspace}
\usepackage{url}
\usepackage{tikz}
\usepackage{fontawesome}
\usepackage{lscape}
\usepackage{color}
\usepackage{anyfontsize}
\usepackage{comment}
\usepackage{soul}
\usepackage{gensymb}
\usepackage{multibib}
\usepackage{tcolorbox}
\usepackage{balance}
\usepackage{footmisc}
\usepackage{tcolorbox}
\usepackage{colortbl}

\usepackage[ruled,linesnumbered]{algorithm2e}
\def\BibTeX{{\rm B\kern-.05em{\sc i\kern-.025em b}\kern-.08em
    T\kern-.1667em\lower.7ex\hbox{E}\kern-.125emX}}

\usepackage{xspace}

\definecolor{box-white}{cmyk}{0, 0, 0, 0, 0}

\definecolor{darkblue}{rgb}{0.0,0.0,0.6}

\newboolean{showcomments}
\setboolean{showcomments}{false} 
\ifthenelse{\boolean{showcomments}}
  {\newcommand{\nb}[2]{
  \fbox{\bfseries\sffamily\scriptsize#1}
     {\sf\small$\blacktriangleright$\textit{\textcolor{red}{#2}}$\blacktriangleleft$}
   }
  }
  {\newcommand{\nb}[2]{}
   
  }

\newcommand{\DC}{\textsc{DeepCrime}\@\xspace}

\newcommand{\AutoT}{\textsc{AutoTrainer}\@\xspace}
\newcommand{\dfd}{\textsc{DeepFD}\xspace}
\newcommand{\DL}{\textsc{DeepLocalize}\xspace}
\newcommand{\DD}{\textsc{DeepDiagnosis}\xspace}
\newcommand{\UM}{\textsc{UMLAUT}\xspace}
\newcommand{\AT}{\textsc{AutoTrainer}\xspace}
\newcommand{\NL}{\textsc{Neuralint}\xspace}

\newcommand{\etal}{\textit{et al.}\xspace}

\newcommand{\COMMENT}[1]{}

\definecolor{codegreen}{RGB}{0, 160, 0}
\definecolor{codered}{RGB}{160, 0, 0}

\begin{document}

\title{An Empirical Study of Fault Localisation Techniques for Deep Learning}

\makeatletter
\newcommand{\linebreakand}{%
  \end{@IEEEauthorhalign}
  \hfill\mbox{}\par
  \mbox{}\hfill\begin{@IEEEauthorhalign}
}
\makeatother

\author{Nargiz Humbatova\\
	Software Institute\\
	Università della Svizzera italiana (USI)\\
	Lugano, Switzerland \\
	\texttt{nargiz.humbatova@usi.ch} \\
	\And
	Jinhan Kim \\
	Software Institute\\
	Università della Svizzera italiana (USI)\\
	Lugano, Switzerland \\
	\texttt{jinhan.kim@usi.ch} \\
	\AND
	Gunel Jahangirova \\
	Department of Informatics \\
	King's College London \\
	London, UK \\
	\texttt{gunel.jahangirova@kcl.ac.uk} \\
	\And
	Paolo Tonella \\
	Software Institute\\
	Università della Svizzera italiana (USI)\\
	Lugano, Switzerland \\
	 \texttt{paolo.tonella@usi.ch} \\
	 \And
	Shin Yoo \\
	School of Computing\\
	KAIST \\
	Daejeon, Republic of Korea \\
	\texttt{shin.yoo@kaist.ac.kr} \\
}

\maketitle

\begin{abstract}
With the increased popularity of Deep  Neural Networks (DNNs), increases also the need for tools to assist developers in the 
DNN implementation, testing and debugging process. Several approaches have been proposed
that automatically analyse and localise potential faults in DNNs under test.
In this work, we evaluate and compare existing state-of-the-art fault localisation techniques, which operate based on both dynamic and static analysis of the DNN. The evaluation is performed on a benchmark consisting of both real faults obtained from bug reporting platforms and faulty models produced by a mutation tool. Our findings indicate that the usage of a single, specific ground truth (e.g., the human defined one) for the evaluation of
DNN fault localisation tools results in pretty low performance (maximum average recall of 0.31 and precision of 0.23). However, such figures increase when considering alternative, equivalent patches that exist for a given faulty DNN. Results indicate that \dfd is the most effective tool, achieving an average recall of 0.61 and precision of 0.41 on our benchmark.

\end{abstract}

\keywords{deep learning \and real faults \and fault localisation}

\section{Introduction}
\label{sec:intro}

Fault localisation (FL) for DNNs is a rapidly evolving area of DL testing~\cite{deeplocalize, wardat2022deepdiagnosis, deepfd,
nikanjam2021automatic, schoop2021umlaut}. The decision logic of traditional software systems is encoded in their source code. Correspondingly, fault localisation for such systems consists of identifying the parts of code that are most likely responsible for the encountered misbehaviour. Unlike traditional software systems, however, the decision logic of DL systems depends on many components such as the model structure, selected hyper-parameters, training dataset, and the framework used to perform the training process. Moreover, DL systems are stochastic in nature, as a retraining with the exactly same parameters might lead to a slightly different final model and performance. These distinctive characteristics make the mapping of a misbehaviour (e.g., poor classification accuracy) to a specific fault type a highly challenging task.

Existing state-of-the-art works~\cite{deeplocalize,autotrainer,wardat2022deepdiagnosis,bakerdetect,nikanjam2021automatic} that focus on the problem of fault localisation for DL systems were shown to be adequate for this task when evaluated on different sets of real-world problems extracted from StackOverflow and GitHub platforms or were deemed useful by developers in the process of fault localisation and fixing~\cite{schoop2021umlaut}. However, these approaches rely on patterns of inefficient model structure design, as well as a set of predefined rules about the values of internal indicators measured during the DL training process. This makes the effectiveness of these approaches highly dependent on the identified set of rules and on the threshold values selected to discriminate the values of the internal indicators of a fault.

To understand whether these tools effectively generalise to a diverse set of fault types and DL systems, and thus, are effective for the real-world usage, we performed an empirical evaluation on a curated benchmark of carefully selected subjects. In this benchmark, the faults obtained by the artificial injection of defects into otherwise well-performing DL models are combined with a set of reproduced real-world DL faults. We ensured that our evaluation involves models of different structure and complexity that solve problems from different domains. The existing evaluations of FL tools are performed only on datasets in which there is a single ground truth repair for each fault. However, the improvement in the performance of a DL model can be achieved by applying different but equally effective fixes. Therefore, limiting the evaluation to a single ground truth might lead to correctly suggested alternative fixes being classified as incorrect, which posits a significant threat to the validity of the performed experiments. To address this issue, we perform a \textit{neutrality analysis}, which aims to identify multiple alternative patches that fix a fault in the DL model. We then evaluate the FL tools considering not only the manually identified single ground truth, but all the available fixes.

Our results show that existing DNN FL techniques produce stable results in a relatively small amount of time ranging from an average of 9 to 278 seconds. The accuracy of fault localisation techniques with regards to the ground truth provided for each issue, however, is quite low (with maximum average recall of 0.31 and precision of 0.23). Once we extend the available ground truth to the changes that produce equivalent or superior improvement in the model's performance, the fault localisation performance significantly improves (the observed maximum recall growth is from 0.31 to 0.61). This demonstrates that the evaluation of the approaches on just one variant of the ground truth does not indeed provide accurate results.

Results indicate that the highest FL performance are achieved by \dfd, which is also the tool that requires the longest execution  time. \NL is extremely efficient, as it relies on static analysis and it requires no model training, although its performance are inferior to those of \dfd.

Overall, we make the following contributions in this paper:
\begin{itemize}
\item An empirical evaluation of four state-of-the-art fault localisation tools on a set of real-world and artificially injected faults, in terms of fault detection effectiveness, efficiency and stability of the results (both considering a single, fixed ground truth, as well as multiple ground truths obtained by means of neutrality analysis).

\item An augmented dataset of DL programs for fault localisation with an extended ground truth based on neutrality analysis.

\item An analysis of the output messages of fault localisation tools in terms of actionability. 

\end{itemize}

\section{Background}
\label{sec:background}

Most of the proposed approaches for fault localisation for DL systems focus
on analysing the run-time behaviour during  model training. According to the
collected information and some predefined rules, these approaches decide whether
they can spot any abnormalities and report them~\cite{deeplocalize,
wardat2022deepdiagnosis, schoop2021umlaut}.

\textbf{DeepLocalize and DeepDiagnosis.} During the training of a DNN, \DL (DL)~\cite{deeplocalize} collects various performance indicators such as loss values, performance metrics, weights,
gradients, and neuron activation values. The main idea behind this approach is that the historic trends in the performance evaluation or the values propagated between layers can serve as an indicator of a fault's presence. To allow the collection of the necessary data, a developer should insert a custom callback provided by the tool into the source code regulating the training process. A callback is a mechanism that is invoked at different stages of  model training (e.g., at the start or end of an epoch, before or after a single batch~\cite{keras}) to perform a set of desired actions -- store historical data reflecting the dynamics of the training process in our case. The tool then compares the analysed values with a list of pre-defined failure symptoms and root causes, which the authors have collected from the existing literature. Based on the performed analysis, \DL either claims that the model is \textit{correct} or outputs an error message listing the detected misbehaviours. The final output of \DL contains the (1) fault type, (2) the layer and the phase (feed forward or backward propagation) in which the DL program has a problem, and (3) the iteration in which learning is stopped. The faults that the tool is able to detect include the following: "Error Before/After Activation", "Error in Loss Function", "Error Backward in Weight/$\Delta$ Weight", and "Model Does Not Learn" that suggests an incorrectly selected learning rate.

\DD (DD)~\cite{wardat2022deepdiagnosis} was built on the basis of \DL and improved the latter by enlarging the list of detected symptoms and connecting them to a set of actionable suggestions. It detects ten types of faults: "Numerical Errors", "Exploding Tensor", "Unchanged Weight", "Saturated Activation", "Dead Node", "Activation Function's Output Out of Range", "Loss Not Decreasing", "Invalid Loss", "Invalid Accuracy", "Accuracy Not Increasing", and "Vanishing Gradient". Depending on the symptom, the actionable messages provided by \DD suggest to change either the loss or optimisation function, layer number, initialisation of weights, learning rate, or indicating that training data is "improper". The authors perform an empirical evaluation of \DD and compare it to \DL, \UM and \AT. They report the time overhead caused by each tool with respect to the number of faults it can successfully localise. However, the time overhead is not reported for each fault but for each dataset (i.e., for all faults of subject MNIST or Circle). In their empirical evaluation, the authors take the randomness associated with model training into account by running each of the compared tools 5 times. 

As \DL does not provide an output that can be translated into a specific fault affecting the model, we only use \DD in the empirical comparison of fault localisation tools. 

\textbf{UMLAUT.} Similarly to \DL and \DD, \UM (UM)~\cite{schoop2021umlaut} operates through a callback attached to the model training call. This tool combines dynamic monitoring of the model behaviour during  training with heuristic static checks of the model structure and its parameters. As an output, it provides the results of the checks along with best practices and suggestions on how to deal with the detected faults. The tool comprises ten heuristics for which the authors found mentions in different sources, such as API documentation and existing literature, lecture notes and textbooks, courses, blogs and non-scientific articles. The heuristics are divided by the area of application into "Data Preparation", "Model Architecture" and "Parameter Tuning". "Data Preparation" heuristics are dynamic and check if the training data contains "NaN", has invalid shape, is not normalised or if the validation accuracy is higher than 95\% after the third epoch of the training. On the other hand, all "Model Architecture" heuristics are static and are focused on the usage of correct activation functions. The "Parameter Tuning" category combines both dynamic and static rules that aim to detect over-fitting and control the values of the learning and dropout rates. As an output, the tool returns a list of heuristics that were violated. The empirical evaluation of \UM was performed with 15 human participants and aimed mostly to determine whether it is useful for the developers. The authors  considered only 6 bugs artificially injected across two DL systems. The reported results do not include the time the tool takes to run. Moreover, the authors do not account for the randomness associated with using the tool and do not compare it to any state-of-the-art tools.

\textbf{Neuralint.} Nikanjam \etal~\cite{nikanjam2021automatic} propose \NL (NL), a model-based fault detection approach that uses meta-modelling and graph transformations. The technique starts with building a meta-model for DL programs that consists of their base structure and fundamental properties. It then performs a verification process that applies 23 pre-defined rules, implemented as graph transformations, to the meta-model, to check for any potential inefficiencies. The rules are classified into four high-level root causes as suggested by Zhang~\etal~\cite{Zhang:2018}:
"Incorrect Model Parameter or Structure" (five rules), "Unaligned Tensor" (four rules), "API Misuse" (five rules), and "Structure Inefficiency" (nine rules). One example of "Unaligned Tensor" rule is a check whether consecutive layers in a model are compatible or whether the reshape of data did not lead to the loss of any elements. "API Misuse" includes a rule to check if the optimiser is correctly defined and connected to the computational graph.
Another rule in this category inspects the parameter initialisation to detect the cases when initialisation is performed more than once or after the training has already started. The "Incorrect Model Parameter or Structure" rule checks if weights and biases are initialised with appropriate values and if suitable activation functions are used for specific layer types. "Structure Inefficiency" is responsible for detecting flaws in the design and structure of DNN that can result in the drop of model performance. Among others, there are rules in this category that 
check if the number of neurons in fully connected layers is decreasing when moving from the input to the output layer or rules that check that pooling layers are not used after each applied convolution, to avoid losing too much information about an input. The empirical evaluation of \NL does not include a comparison to any of the existing tools. The authors also do not report the time it takes to run the tool for each of the faults, but only provide information on the time for 5 selected DL models with different sizes.

\textbf{DeepFD.} \dfd~\cite{deepfd} (DFD) is a learning-based framework that leverages mutation testing and popular ML algorithms to construct a tool capable of labelling a given DL program as correct or faulty according to a list of common fault types the tool has learned to detect. To train the classifiers that lie at the core of the technique, the authors prepare a set of correct and faulty models. Faulty models are obtained through the artificial injection of up to five mutations to each correct program being used. The mutations that are used to inject faults are changing loss or optimisation function, changing learning rate, decreasing number of epochs, and changing activation functions. Consequently, these fault types correspond to the fault localisation capabilities of the tool. To construct the training dataset, all of the generated mutants and the original models are trained, while collecting run-time data of the same kind as for \DL, \DD, \UM. The authors then extract 160 features from these data using statistical operations (e.g., calculating skewness, variance or standard deviation). As the next step, three popular ML algorithms (K-Nearest Neighbors~\cite{knearestn}, Decision Tree~\cite{breiman2017classification} and Random Forest~\cite{ho1995random}) are trained on the created dataset. A union of the prediction results of these classifiers is used for fault localisation in a given program under test. 
\dfd outputs a list of detected faults along with the code lines affected by each fault type. The empirical evaluation of \dfd contains comparison to \AT and \DL. The authors take the stochasticity of the proposed approach into account and run their experiments 10 times. However, no information is reported on the time required to train and run \dfd. 

\textbf{Autotrainer.} 
\AutoT \cite{autotrainer} is an automated tool whose aim is to detect, localise, and repair training related problems in DL models. It starts with an already trained and saved underperforming DL model. To  check the model, \AutoT continues the training process and observes different internal parameters such as loss values, accuracy, gradients, etc. The collected information is then analysed and verified against a set of rules that are aimed to detect potential training failures. In particular, the authors focus on the symptoms of exploding and vanishing gradients, oscillating loss, slow convergence and `dying ReLU'. To deal with each of the reveled symptom, \AutoT applies an ordered set of possible solutions to the model. After a potential solution is applied, the tool continues training the model for one more epoch to check whether the symptom is gone or not. The solutions include changing activation functions, hyperparameters such as learning rate and batch size, optimisers and weight initialisation, and addition of batch normalisation layers to the model structure. We do not consider \AutoT in this empirical comparison as its final goal is to patch an already trained model rather than localise and fix the source of a DNN's misbehaviour.

\textbf{Fault types covered by the tools.} 
In Table~\ref{tab:abbr} we introduce abbreviations for the fault types that affect issues from our benchmark (see Table~\ref{tab:benchmark}) and those that are suggested in the output of the evaluated fault localisation tools. Most of the abbreviations (except for the last two) are adopted from the mutation operators of DeepCrime~\cite{deepcrime} that were designed to inject the corresponding real faults into DL programs. Sometimes the abbreviation is followed by a list of layer identifiers within brackets, to indicate which layers are affected by a fault. For instance, ACH(1, 3) means that the activation function should be changed in layers 1, 3.

\begin{table*}[htb]
    \caption{Fault types and abbreviations}
    
    \begin{small}
    \begin{center}
    \scalebox{0.9}{
    \begin{tabular}{cl}
    \toprule
    Abbreviation & Fault Type \\ 
    \midrule
    HBS & Wrong batch size\\
    HLR & Wrong learning rate \\
    HNE & Change number of epochs \\
    ACH & Change activation function  \\
    RAW & Redundant weights regularisation \\
    WCI & Wrong weights initialisation \\
    LCH & Wrong loss function \\
    OCH & Wrong optimisation function \\
    LRM & Missing layer \\
    LAD & Redundant layer \\
    LCN & Wrong number of neurons in a layer \\
    LCF & Wrong filter size in a convolutional layer \\
    BCI & Wrong bias initialisation \\
    CPP & Wrong data preprocessing \\
    \bottomrule
    \end{tabular}
    }
    \end{center}
    \end{small}
    
    \label{tab:abbr}
    \end{table*}
    
As was noted before, authors of \dfd identified five more frequent types of faults in the benchmark they used to evaluate their tool and  designed their tool to detect these specific fault types. In particular, they cover five fault types: HLR, ACH, LCH, HNE and OCH (a description of the considered fault types is provided later, in Table~\ref{tab:abbr}),  which despite being a small number of all the recognised faults that affect DL systems, are indeed quite frequently experienced in the real world~\cite{taxonomy}. \DD, while observing the training process and the changes it brings to the internal variables of a DNN, is designed to detect 10 different fault symptoms, such as vanishing gradients or numerical errors. It maps these symptoms to 7 different fault types. Similarly to \dfd, they can detect LCH, ACH, HLR, OCH faults, and additionally, the tool can detect the WCI fault, as well as problems in the training data and in the number of layers. \UM operates based on a set of both dynamic and static checks performed before and during training of a model. The used heuristics cover hyperparameter tuning, and while problems with learning rate are detected also by the previously discussed approaches, \UM can also detect high drop out rate, while it does not pay attention to the number of epochs as \dfd~\cite{schoop2021umlaut}. Similarly to \DD, \UM can detect problems with training data and its pre-processing and also covers problems with activation functions of the model. \NL relies on meta-modelling and graph transformations to perform a verification process based on 23 pre-defined rules that cover initialisation of different parameters, nuances of neural network architecture, and API misuse~\cite{nikanjam2021automatic}. The detection of the violation of these rules leads to a number of diverse recommended fixes that cover data pre-processing, selection of optimiser, activation functions and tuning of the neural network architecture. Despite some similarities in the types of detected faults, all of the approaches have their own specifics and vary in the localisation methods used. At the current state of the art, it appears that the available approaches are rather complementary to each other.

The tools described in this section are built using a limited set of rules and best practices, fixed thresholds or training data, resulting in an urgent need to empirically investigate the generalisability of these approaches to diverse programs and architectures. While some of the experiments conducted by the proponents to evaluate these tools include comparisons to other existing tools, no work considers the full set of four existing FL approaches. Moreover, the existing evaluations do not always consider the randomness associated with the training of the DL models and do not report detailed information on the runtime costs associated with the FL tools. Most importantly, no third party evaluation of these tools on a curated dataset of faulty DL models was ever conducted and reported so far.

\section{Neutrality Analysis}

\begin{algorithm}[ht]
  \small
  \SetCommentSty{mycommfont}
  \SetKwInput{KwData}{Input}
  \SetKwInput{KwResult}{Output}

  \KwData{Initial model $s$, GT accuracy $acc_{gt}$, stopping conditions $SC$, and $top_k$}
  \KwResult{Edges $E$ and alternative GTs $R$}
    $Q, Visited, R \leftarrow \emptyset, \emptyset, \emptyset$\\
    $acc_{s} \leftarrow trainAndEvaluate(s)$\\
    $Q.enqueue([s, acc_{s}])$\\
    
    \While{$Q \neq \emptyset$ and $SC$ not met}{
      $c, acc_{c} \leftarrow Q.dequeue()$\\
      $Visited.append(c)$\\
      
      $N \leftarrow getNeighbours(c, Visited)$\\
      $tempQ \leftarrow \emptyset$ \\
      \ForEach{$n$ in $N$}{
        $acc_{n} \leftarrow trainAndEvaluate(n)$\\

        \tcp{Check whether the neighbour is equivalent to or better than the current node.}
        \If{$isNeutral(acc_{n}, acc_{c})$}{
          $tempQ.append([n, acc_{n}])$\\
        }
        \tcp{Check whether the neighbour is equivalent to or better than the given GT.}
        \If{$isNeutral(acc_{n}, acc_{gt})$}{
          $R.append([n, acc_{n}])$\\
        }
      }
      \tcp{Enqueue top $k$ neighbours to $Q$ and make edges to them.}
      $tempQ \leftarrow sort(tempQ, top_k)$\\
      \ForEach{$n, acc_{n}$ in $tempQ$}{
         $Q.enqueue([n, acc_{n}])$\\
         $E.append(c, n)$\\  
      }
    }
    
    \Return{$E, R$}\\

  \caption{Breadth-First Search (BFS) for Neutrality Analysis}
  \label{algo:bfs}
\end{algorithm}

In the context of fault localisation in DNNs, our goal is to identify and localise faults in the network architecture and hyperparameters. Even though the benchmark of faulty models includes repaired and correct versions of DL models, there are no strict rules and guidelines as to which combinations and architectures would perform better for specific tasks. Therefore, we posit that there exist potential for finding alternative patches that not only supplement the known patch by suggesting different ways of repairing but also possibly exhibit better performance than the known one. Identifying such alternative patches would enable a more precise evaluation of FL techniques.

In our search for alternative patches, we are inspired by the notion of software neutrality, which states that a random mutation to an executable program is considered \emph{neutral} if the behaviour of the program on the test set does not change~\cite{neutrality}. This neutrality analysis aims to investigate diverse patches with similar or better fitness: these can be utilised as alternative Ground Truths (GTs). Since our targets are DL programs, the conditions for performing neutrality analysis differ from those of traditional programs. For example, the fitness is now measured by the model performance with  standard metrics such as test set accuracy. This means that fitness evaluation involves training and testing of the model. Moreover, during fitness evaluation it is important to account for the inherent stochastic properties of DL programs because the model's performance can vary with multiple trainings. To address this, in our algorithm below, we train the model ten times and calculate the fitness as an average of the resulting ten accuracy values.

\begin{table*}[!t]
    \caption{Mutation Operator for Neutrality Analysis}
    \centering
    \scalebox{0.95}{
    \begin{tabular}{ll}
    \toprule
    Operator & Description \\ 
    \midrule
    
    Change activation function & \multirow{5}{*}{\makecell[l]{It replaces the value of the hyperparameter of the given \\model with other pre-defined values from Keras.}} \\
    Change kernel initialisers &  \\
    Change bias initialisers & \\
    Change loss function & \\
    Change optimiser & \\
    \midrule
    Change learning rate & \makecell[l]{It changes the learning rate by either multiplying it by\\ ten or dividing it by ten.} \\
    \midrule
    Change epochs & \makecell[l]{It changes the epochs by either multiplying it by two or\\ dividing it by two.} \\ 
    \midrule
    Change batch size & \makecell[l]{It replaces the batch size with other pre-defined values \\of 16, 32, 64, 128, 256, 512.} \\
    \midrule
    Change layer & \makecell[l]{It either duplicates or deletes the given layer.} \\ 
    \midrule
    \makecell[l]{Change the number of neurons \\in a given dense layer} & \makecell[l]{It changes the number of neurons by either multiplying \\its number by two or dividing it by two.} \\
    
    \bottomrule
    \end{tabular}
    }
    
    \label{tab:mut_ops}
\end{table*}

Algorithm~\ref{algo:bfs} presents the Breath-First Search (BFS) for our neutrality analysis on DL programs. This algorithm takes as inputs an initial (buggy) model $s$, the accuracy of the known GT $acc_{gt}$, and stopping criteria $SC$. The outputs are a list of alternative GTs and edges of the neutrality graph. The algorithm starts with training and evaluating the initial buggy model before putting it in the queue (Lines 2-3). Next, it begins a search loop where it iteratively retrieves a model (i.e., a parent model $c$) along with its accuracy $acc_{c}$ from the queue (Line 5). Subsequently, the algorithm explores all adjacent models (i.e., neighbours) that are obtained by applying a distinct single mutation on $c$ (Line 7). Each mutation involves changing a single hyperparameter of the model, in other words, neighbouring models differ from their parent model by one hyperparameter. The details of mutation operators adopted from Kim \etal~\cite{emp_repair} are shown in Table~\ref{tab:mut_ops}.
Then, the algorithm iterates over the neighbours by training and evaluating them (Line 10), and evaluates the \textit{neutrality}\footnote{A model is considered \textit{neutral} relative to another model if it has equivalent or higher fitness than the other's, by comparing the mean accuracy of ten trained instances of the model and the other model.} of each neighbour compared to the parent model (Line 11) and the known GT (Line 13). 
Since sometimes the number of neutral neighbours is numerous, potentially impeding the exploitation of the search, the algorithm places them into the temporal queue, not in the main queue (Line 12). If it is neutral with respect to the known GT, it is added to a list of alternative GTs (Line 14). After this iteration, the algorithm sorts the temporal queue by accuracy and takes only top-k performing neighbours by enqueueing them into the main queue. The search process stops when it meets the given stopping criteria or the queue is empty. As the algorithm evolves the model by applying mutation to its parent, the resulting alternate GTs are usually higher-order mutants of the initial buggy model.

\begin{figure}[!t]
    \centering
    \includegraphics[width=0.7\linewidth]{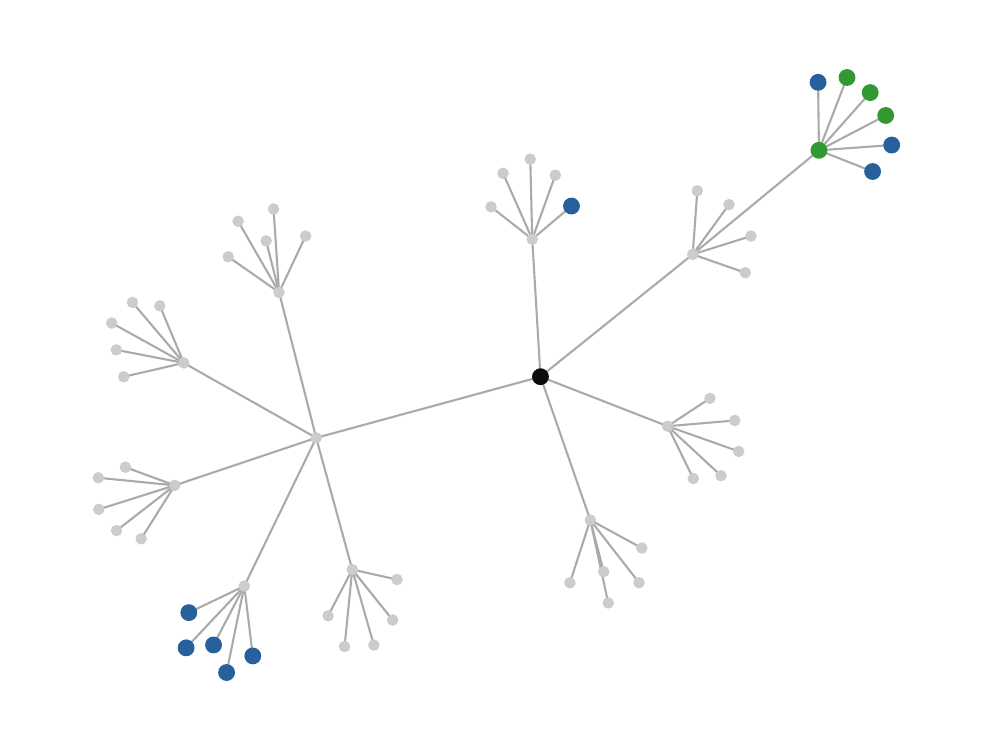}
    \caption{An example neutrality network of D4}
    \label{fig:network}
\end{figure}

Based on the search results, we can draw a so-called neutrality graph or `network'~\cite{neutrality}, as shown in Figure~\ref{fig:network}. Each edge represents a single mutation and each node represents the DL models (i.e., mutants). A black node denotes the initial buggy model and other nodes are neutral with their parent node. Among them, the ones that are on par with or better than the known patch are coloured in either blue or green. In particular, models that outperform the known GT with statistical significance are marked blue.\footnote{For the computation of statistical significance, we employ a Generalized Linear Model (GLM) with a significance level 0.05 and Cohen's $d$ to measure the effect size, for which we adopt a threshold at 0.5.} Those that are not statistically significant but exceed the known patch in terms of average accuracy are coloured in green. In this example, we found 13 alternative patches that fix the buggy model differently but show equal or higher performance than the known patch.

\section{Empirical study} \label{sec:empirical_study_fl}

\subsection{Research Questions}

The \textit{aim} of this empirical study is to compare existing DL fault localisation approaches and to explore their generalisability to different subjects represented by our benchmark of artificial and real faults. To cover these objectives, we define the following four research questions:

\begin{itemize}
    \item \textbf{RQ1. Effectiveness}: \textit{Can existing FL approaches identify and locate defects correctly in faulty DL models?}
    \item \textbf{RQ2. Stability}: \textit{Is the outcome of fault identification analysis stable across several runs?}
    \item \textbf{RQ3. Efficiency}: \textit{How costly are FL tools when compared to each other?}
\end{itemize}

RQ1 is the key research question for this empirical study, as it compares the effectiveness of different FL tools on our curated benchmark of artificial and real faults. We conduct our effectiveness analysis twice, first with the available ground truth and then with the ground truth extended by neutrality analysis.

RQ2 and RQ3 investigate two important properties of the FL tools being compared: their stability across multiple executions and their execution cost.

\subsection{Benchmark}

To evaluate and compare the fault localisation techniques selected for this study, we adopt a carefully selected benchmark of faulty models from the existing literature~\cite{emp_repair}. This benchmark is of a particular interest as it combines both models affected by real-world faults and those deliberately produced using artificial faults. 

\textit{Artificial faults} of this benchmark were produced by \DC~\cite{deepcrime}, a state-of-the-art mutation testing tool for DL systems based on real faults~\cite{taxonomy}. The models subject to  fault injection by \DC cover a diverse range of application areas, such as handwritten digit classification (MN), speaker recognition (SR), self-driving car designed for the Udacity simulator (UD), eye gaze prediction (UE), image recognition (CF10), and news categorisation (RT). This section of the benchmark contains 21 faulty models generated by injecting 9 distinct fault types into originally well-performing models. As \DD was not applicable to SR, UD, and UE, we had to limit our evaluation to the remaining subjects.

The \textit{real faults} section of the benchmark was derived from the set of issues that were collected and used for the evaluation of the \dfd tool~\cite{deepfd,deepfd_replication}. This initial set contains 58 faulty DNNs collected from bug-reporting platforms such as StackOverflow and Github. 

Later, Kim et al.~\cite{emp_repair} curated this benchmark by performing a series of checks and filtered out those issues that were not in fact correctly reproduced from the corresponding fault report. Specifically, first of all they checked if there was a match between the model, dataset, and fix in the code and the post from SO and GitHub. As  second step, they have discarded the issues where the fault was not exposed in the buggy version of the code when the program was run or was not actually repaired in the fixed program. As a result, only 9 out of 58 issues  proved to be reliably reproducible. We adopt this benchmark of 9 real faults in our empirical comparison of FL tools.

The final benchmark that was used in this study can be found in Table \ref{tab:benchmark}. Column `Fault Type' shows whether the fault was real (`R') or artificially seeded (`A'); column `Id' bears the ID of the fault which will be reused throughout the paper; it  refers to real faults curated from the \textsc{DeepFD} dataset (prefix D), MNIST mutants (M), CIFAR mutants (C), Reuters mutants(R); column `SO Post \#/Subject' provides the StackOverflow post number from which the fault was obtained in the case of real faults and the subject name in the case of artificial faults; column `Task' has `C' for faults that solve a classification problem and `R' for those dealing with a regression task; column `Faults' contains the available Ground Truth (GT), i.e., the list of faults known to affect each model (before neutrality analysis).

\begin{table*}[htb]
    \caption{Benchmark: the Fault Type can be real (R) or artificial (A); the fault Id identifies real faults curated from the \textsc{DeepFD} dataset (D), MNIST mutants (M), CIFAR mutants (C), Reuthers mutants(R); the task can be classification (C) or regression (R).}
    \begin{small}
    \begin{center}
    \scalebox{0.8}{
    \begin{tabular}{ccccl}
    
    \toprule
    
    \multirow{2}{*}{Fault Type} & \multirow{2}{*}{Id} & SO Post \# & \multirow{2}{*}{Task} & \multirow{2}{*}{Faults}\\
    
    & & /Subject & &  \\
    
    \midrule
    R & D1 & 31880720 & C & Wrong activation function \\
    
    \midrule
    \multirow{2}{*}{R} & \multirow{2}{*}{D2} & \multirow{2}{*}{41600519} & \multirow{2}{*}{C} & Wrong optimiser | Wrong batch size \\
    & & & & Wrong number of epochs \\
    
    \midrule
    \multirow{3}{*}{R} & \multirow{3}{*}{D3} &  \multirow{3}{*}{45442843} & \multirow{3}{*}{C} & Wrong optimiser | Wrong loss function \\
    & & & &  Wrong batch size |  Wrong activation function \\
    & & & & Wrong number of epochs  \\
    
    \midrule
    \multirow{2}{*}{R} & \multirow{2}{*}{D4} & \multirow{2}{*}{48385830} & \multirow{2}{*}{C} & Wrong activation function  \\
    & & & & Wrong loss function | Wrong learning rate \\
    
    \midrule
    R & D5 & 48594888 & C & Wrong number of epochs | Wrong batch size  \\
    
    \midrule
    \multirow{2}{*}{R} & \multirow{2}{*}{D6} & \multirow{2}{*}{50306988} & \multirow{2}{*}{C} & Wrong learning rate | Wrong number of epochs \\
    & & & & Wrong loss function | Wrong activation function\\
    
    \midrule 
    R & D7 & 51181393 & R & Wrong learning rate \\
    
    \midrule
    R & D8 & 56380303 & C & Wrong optimiser | Wrong learning rate \\
    
    \midrule
    \multirow{2}{*}{R} & \multirow{2}{*}{D9} & \multirow{2}{*}{59325381} & \multirow{2}{*}{C} & Wrong preprocessing  \\
    & & & & Wrong activation function | Wrong batch size 
    \\

    \midrule
    A & M1 & MN & C & Wrong weights initialisation \\
    \midrule
    A & M2 & MN & C & Wrong activation function \\
    \midrule
    A & M3 & MN & C & Wrong learning rate \\
    \midrule
    A & C1 & CF10 & C & Wrong activation function \\
    \midrule
    A & C2 & CF10 & C & Wrong number of epochs \\
    \midrule
    A & C3 & CF10 & C & Wrong weights initialisation \\
    \midrule
    A & R1 & RT & C & Wrong weights regularisation\\
    \midrule
    A & R2 & RT & C & Wrong activation function \\
    \midrule
    A & R3 & RT & C & Wrong learning rate \\
    \midrule
    A & R4 & RT & C & Wrong loss function \\
    \midrule
    A & R5 & RT & C & Wrong optimisation function\\
    \midrule
    A & R6 & RT & C & Wrong weights initialisation \\
    \midrule
    A & R7 & RT & C & Wrong activation function \\

    \bottomrule
    \end{tabular}}
    \end{center}
    \end{small}
    \label{tab:benchmark}
    \end{table*}

\subsection{Experimental Settings \& Evaluation Metrics} \label{sec:exp_fl}
For the comparison, we adopt publicly available versions of all considered tools~\cite{deepfd_replication, umlaut_replication, neuralint_replication, deepdiagnosis_replication} that are run on Python with library versions specified in the requirements for each tool. However, we had to limit the artificial faults to those obtained using CF10, MN, and RT as \DD is not applicable to other subjects.

The authors of \dfd adopted the notion of statistical mutation killing~\cite{JahangirovaICST20} in their tool. They run each of the models used to train the classifier as well as the model under test 20 times to collect the run-time features. For  fault localisation using \dfd, we adopt an ensemble of already trained classifiers provided in the tool's replication package. Similar to the authors, for each faulty model in our benchmark, we collect the run-time behavioural features from 20 retrainings of the model. \NL is based on static checks that do not require any training and thus, are not prone to randomness. We run each of the remaining tools 20 times to account for the randomness in the training process and report the most frequently observed result (mode).

To calculate the similarity between the ground truth provided for each fault in our benchmark and the fault localisation results, we adopt the standard information retrieval metrics Precision (PR), Recall (RC) and $F_{\beta}$ score:

\begin{equation}
    RC = \frac{| FT_{loc} \cap FT_{gt} |}{| FT_{gt} |}
\end{equation}

\begin{equation}
    PR = \frac{| FT_{loc} \cap FT_{gt} |}{| FT_{loc} |}
\end{equation}

\begin{equation}
    F_{\beta} = (1 + \beta^2) \frac{PR \cdot RC}{\beta^2 PR + RC}
\end{equation}

Recall measures the proportion of correctly reported fault types in the list of localised faults ($FT_{loc}$) among those in the ground truth ($FT_{gt}$); Precision measures the proportion of correctly reported fault types among the localised ones; $F_{\beta}$ is a weighted geometric average of $PR$ and $RC$, with the weight $\beta$ deciding on the relative importance between $RC$ and $PR$. Specifically, we adopt $F_{\beta}$ with $\beta$ equals 3, which gives three times more importance to recall than to precision. This choice of beta is based on the assumption that in the task of fault localisation, the ability of the tool to find as many correct fault sources as possible is more important than the precision of the answer.

For neutrality analysis, we set $top_k$ to 5 and the stopping condition $SC$ to a 48-hour time budget. During the search, every model is trained ten times and we use a mean of the ten metric values depending on the task solved by each subjects (i.e., accuracy for classification or loss for regression).

\section{Results}
\label{sec:results}

\subsection{RQ1 (Effectiveness before neutrality analysis)}

Tables~\ref{tab:main_results_fl_1_4_1}, \ref{tab:main_results_fl_1_4_2}, \ref{tab:main_results_fl_1_4_3}, and \ref{tab:main_results_fl_1_4_4} present the output of the application of fault localisation tools (DFD, DD, NL, and UM, respectively) to our benchmark. Column \textit{`GT'} stands for \textit{`Ground Truth'} and provides the list of fault types affecting the model, while  column `\text{\#F}' reports the length of this list. Column \textit{`<tool\_name>-output'} contains the fault list generated by each FL tool, while column \textit{`Matches-GT'} indicates for each fault in the ground truth whether it was detected by the tool or not (1 if yes and 0 otherwise) and column \textit{`\#M'} counts the number of detected faults. For each row (issue) we underline the number of detected faults (`$\#M$') if the tool was able to achieve the best result across all the compared approaches. We provide the average number of fault types detected  for issues generated by artificially injected faults or real-world issues (rows `Avg.') and across the whole benchmark (row `T.A.', i.e., Total Average).

When faults affect only selected layers, we specify the indexes of the faulty layers within brackets, for ground truth and for fault localisation results, if this information is provided. Moreover, `-' means that an FL tool was not able to find any fault in the model under test; `N/A' means that the tool was not applicable to the fault type in question or crashed on it. For example, \NL accepts only optimisers that are defined as strings (e.g., `sgd'), which in turn implies that the default learning rate as defined by the framework is used. This makes it not possible for \NL to find an optimiser with modified learning rate. Symbol `,' separates all detected faults, while separation by `|' means that the faults are alternative to each other, i.e., the tool suggests either of them could be the possible cause of model's misbehaviour.

\begin{table*}[htb]
  \centering
  \caption{Ground Truth (GT) and fault localisation outcome generated by \dfd (DFD); \#F indicates the number of detected faults, while \#M the number of matched faults (with underline used to indicate the best result among all tools being compared). Avg. shows the average within artificial or real faults. T.A. shows the total average across faults.} 
  \label{tab:main_results_fl_1_4_1}
   \scalebox{0.82}{
  \begin{tabular}{l|l|c|l|c|l}
    \toprule
    Id & GT  &\#F& Matches-GT  &\#M& DFD-output \\
    \midrule    
M1 & WCI(0)  &1& 0  &0 & HLR, ACH, LCH, HNE  \\
M2 & ACH(7)  &1& 0  &0 & OCH, HLR, HNE  \\
M3 & HLR  &1& 1  &\underline{1} & OCH, HLR, LCH  \\
C1 & ACH(2)  &1& 1  &\underline{1} & OCH, HLR, ACH, LCH  \\
C2 & HNE  &1& 0  &0 & OCH, ACH, LCH  \\
C3 & WCI(2)  &1& 0  &0 & OCH, ACH, LCH, HNE  \\
R1 & RAW(0)  &1& 0  &0 & HLR, LCH, HNE  \\
R2 & ACH(2)  &1& 0  &0 & OCH, LCH, HNE  \\
R3 & HLR  &1& 0  &0 & OCH, LCH, HNE  \\
R4 & LCH  &1& 1  &\underline{1} & ACH, LCH  \\
R5 & OCH  &1& 1  &\underline{1} & OCH, ACH, HNE  \\
R6 & WCI(0)  &1& 0  &0 & OCH, ACH, LCH, HNE  \\
R7 & ACH(2)  &1& 0  &0 & OCH, LCH, HNE  \\
\textbf{Avg.}& & \textbf{1}& & \textbf{0.3}& \\
    \midrule
D1 & ACH(7)  &1& 1  &\underline{1} & ACH  \\
D2 & OCH, HNE,  HBS  &3& 0, 0, 0  & 0 & ACH  \\
D3 & OCH, LCH, ACH(0,1), HNE, HBS  & 5 & 1, 0, 0, 0, 0  &1 & OCH, HLR  \\
D4 & ACH(0,1), LCH, HLR  & 3 & 0, 0, 0  & 0 & OCH  \\
D5 & HNE, HBS  & 2 & 0, 0  & 0 & OCH, ACH  \\
D6 & HLR, HNE, LCH, ACH(1)  &4& 1, 1, 0, 0  &\underline{2} & OCH, HLR , HNE  \\
D7 & HLR  &1& 0  &0 & LCH  \\
D8 & OCH, HLR  & 2 & 1, 1  &\underline{2} & OCH, HLR, LCH, HNE  \\
D9 & CPP, ACH(5,6), HBS  &3& 0, 0, 0  & 0 & N/A  \\
 \textbf{Avg.}& & \textbf{2.7}& & \textbf{0.7}&\\
     \midrule
 \textbf{T.A.}& & \textbf{1.7}& & \textbf{0.5}&\\
\bottomrule
  \end{tabular}
  }
\end{table*}

\begin{table*}[htb]
  \centering
  \caption{Ground Truth (GT) and fault localisation outcome generated by \DD (DD); \#F indicates the number of detected faults, while \#M the number of matched faults (with underline used to indicate the best result among all tools being compared). Avg. shows the average within artificial or real faults. T.A. shows the total average across faults.} 
  \label{tab:main_results_fl_1_4_2}
   \scalebox{0.85}{
  \begin{tabular}{l|l|c|l|c|l}
    \toprule
    Id & GT  &\#F& Matches-GT  &\#M & DD-output \\
    \midrule    
M1 & WCI(0)  &1& 0  &0 & HLR  \\
M2 & ACH(7)  &1& 1  &\underline{1} & ACH(7)  \\
M3 & HLR  &1& 0  &0 & -  \\
C1 & ACH(2)  &1& 0  &0 & -  \\
C2 & HNE  &1& 0  &0 & -  \\
C3 & WCI(2)  &1& 0  &0 & -  \\
R1 & RAW(0)  &1& 0  &0 & -  \\
R2 & ACH(2)  &1& 1  &\underline{1} & ACH(2)  \\
R3 & HLR  &1& 0  &0 & -  \\
R4 & LCH  &1& 0  &0 & LRM | LAD | ACH(0)  \\
R5 & OCH  &1& 0  &0 & -  \\
R6 & WCI(0)  &1& 0  &0 & -  \\
R7 & ACH(2)  &1& 1  &\underline{1} & ACH(2)  \\
 \textbf{Avg.}& & \textbf{1}& & \textbf{0.2}&\\
    \midrule
D1 & ACH(7)  &1& 0  &0 & HLR  \\
D2 & OCH, HNE,  HBS  &3& 0, 0, 0  &0 & -  \\
D3 & OCH, LCH, ACH(0,1), HNE, HBS  &5& 0, 0, 0, 0, 0  &0 & -  \\
D4 & ACH(0,1), LCH, HLR  &3& 1, 0, 0  &1 & ACH(1)  \\
D5 & HNE, HBS  &2& 0, 0  &0 & -  \\
D6 & HLR, HNE, LCH, ACH(1)  &4& 0, 0, 0, 0  &0 & -  \\
D7 & HLR  &1& 0  &0 & -  \\
D8 & OCH, HLR  &2& 0, 0  &0 & -  \\
D9 & CPP, ACH(5,6), HBS  &3& 0, 0, 0  &0 & N/A  \\
 \textbf{Avg.}& & \textbf{2.7}& & \textbf{0.1}&\\
 \midrule
 \textbf{T.A.}& & \textbf{1.7}& & \textbf{0.2}&\\
\bottomrule
  \end{tabular}
  }
\end{table*}

\begin{table*}[htb]
  \centering
  \caption{Ground Truth (GT) and fault localisation outcome generated by \NL (NL); \#F indicates the number of detected faults, while \#M the number of matched faults (with underline used to indicate the best result among all tools being compared). Avg. shows the average within artificial or real faults. T.A. shows the total average across faults.} 
  \label{tab:main_results_fl_1_4_3}
   \scalebox{0.8}{
  \begin{tabular}{l|l|c|l|c|l}
    \toprule
    Id & GT  &\#F& Matches-GT  &\#M & NL-output \\
    \midrule    
M1 & WCI(0)  &1& 1  &1& WCI(0)  \\
M2 & ACH(7)  &1& 0  &0 & LCH  \\
M3 & HLR  &1& 0  &0 & N/A  \\
C1 & ACH(2)  &1& 0  &0 & -  \\
C2 & HNE  &1& 0  &0 & -  \\
C3 & WCI(2)  &1& 1  &\underline{1} & WCI(3)  \\
R1 & RAW(0)  &1& 0  &0 & -  \\
R2 & ACH(2)  &1& 0  &0 & LCH  \\
R3 & HLR  &1& 0  &0 & N/A  \\
R4 & LCH  &1& 1  &\underline{1} & LCH  \\
R5 & OCH  &1& 0  &0 & -  \\
R6 & WCI(0)  &1& 1  &\underline{1} & WCI(0)  \\
R7 & ACH(2)  &1& 0  &0 & LCH  \\
 \textbf{Avg.}& & \textbf{1}& & \textbf{0.3}&\\
    \midrule
D1 & ACH(7)  &1& 0  &0 & LCH  \\
D2 & OCH, HNE,  HBS  &3& 0, 0, 0  &0 & -  \\
D3 & OCH, LCH, ACH(0,1), HNE, HBS  &5& 0, 1, 1, 0, 0  &\underline{2} & ACH(1), LCH, LCN(0)  \\
D4 & ACH(0,1), LCH, HLR  &3& 1, 0, 0  &1 & ACH(0), BCI(0,1)  \\
D5 & HNE, HBS  &2& 0, 0  &0 & LCF(0)  \\
D6 & HLR, HNE, LCH, ACH(1)  &4& 0, 0, 0, 0  &0 & -  \\
D7 & HLR  &1& 0  &0 & N/A  \\
D8 & OCH, HLR  &2& 0, 0  &0 & -  \\
D9 & CPP, ACH(5,6), HBS  &3& 0, 0, 0  &0 & ACH(0), LCN(2,3)  \\
 \textbf{Avg.}& & \textbf{2.7}& & \textbf{0.3}&\\
 \midrule
 \textbf{T.A.}& & \textbf{1.7}& & \textbf{0.3}&\\
\bottomrule
  \end{tabular}
  }
\end{table*}

\begin{table*}[htb]
  \centering
  \caption{Ground Truth (GT) and fault localisation outcome generated by \UM (UM); \#F indicates the number of detected faults, while \#M the number of matched faults (with underline used to indicate the best result among all tools being compared). Avg. shows the average within artificial or real faults. T.A. shows the total average across faults.} 
  \label{tab:main_results_fl_1_4_4}
   \scalebox{0.85}{
  \begin{tabular}{l|l|c|l|c|l}
    \toprule
    Id & GT  &\#F& Matches-GT  &\#M& UM-output \\
    \midrule    
M1 & WCI(0)  &1& 0  &0 & HLR  \\
M2 & ACH(7)  &1& 1  &\underline{1}& ACH(7), HLR  \\
M3 & HLR  &1& 0  &0 & -  \\
C1 & ACH(2)  &1& 0  &0 & -  \\
C2 & HNE  &1& 0  &0 & -  \\
C3 & WCI(2)  &1& 0  &0 & -  \\
R1 & RAW(0)  &1& 0  &0 & -  \\
R2 & ACH(2)  &1& 1  &\underline{1} & ACH(2)  \\
R3 & HLR  &1& 1  &\underline{1} & HLR  \\
R4 & LCH  &1& 0  &0 & -  \\
R5 & OCH  &1& 0  &0 & -  \\
R6 & WCI(0)  &1& 0  &0 & -  \\
R7 & ACH(2)  &1& 1  &\underline{1} & ACH(2)  \\
 \textbf{Avg.}& & \textbf{1}& & \textbf{0.3}&\\
    \midrule
D1 & ACH(7)  &1& 0  &0 & -  \\
D2 & OCH, HNE,  HBS  &3& 0  &0 & ACH(7)  \\
D3 & OCH, LCH, ACH(0,1), HNE, HBS  &5& 0, 0, 0, 0, 0  &0 & -  \\
D4 & ACH(0,1), LCH, HLR  &3& 1, 0, 1  &\underline{2} & ACH(0,1), HLR  \\
D5 & HNE, HBS  &2& 0, 0  &0 & -  \\
D6 & HLR, HNE, LCH, ACH(1)  &4& 0, 0, 0, 0  &0 & -  \\
D7 & HLR  &1& 0  &0 & -  \\
D8 & OCH, HLR  &2& 0, 0  &0 & -  \\
D9 & CPP, ACH(5,6), HBS  &3& 0, 0, 0  &0 & ACH(0,2,4)  \\
 \textbf{Avg.}& & \textbf{2.7}& & \textbf{0.2}&\\
 \midrule
 \textbf{T.A.}& & \textbf{1.7}& & \textbf{0.3}&\\
\bottomrule
  \end{tabular}
  }
\end{table*}

Interestingly, in the majority of cases \UM (20 out of 22) and \DD (15 out of 22) suggest changing the activation function of the last layer to `softmax' even if in 73 \% of these cases for \UM and 67\% for \DD, the activation function is already  `softmax'. This also happens once to \NL. We exclude such misleading suggestions from the tools' output. Moreover, sometimes \UM mentions that over-fitting is possible. Since it is just a possibility and such a message does not point to a specific fault, we also exclude it from our analysis. Full output messages provided by the tools are available in our replication package~\cite{fl_comparison_replication}.

Table~\ref{tab:main_results_fl_2} reflects the overall evaluation of the effectiveness of the FL tools. Column `$GT \# F$' shows the number of fault types in the ground truth, while columns \textit{`<tool\_name>'} contain all the measured performance metrics for each tool: columns `$RC$' report the values of Recall, columns `$PR$' report  Precision, and columns `$F_3$' the $F_\beta$ score with $\beta = 3$. We treated the cases when a tool is not applicable to an issue as if the tool has failed to locate any faults affecting the issue. We provide mean values for each tool across artificial and real faults (rows `Avg.') and across all issues in the benchmark (row `T.A'), to ease the comparison between the tools. According to these numbers, \dfd, on average, exhibits the best performance and significantly outperforms other tools on real faults. This can be influenced by the fact that the `Real Fault' part of the benchmark comes from the evaluation benchmark of \dfd, as this was the only available source of truly reproducible real faults. The selection of fault types that \dfd is trained to detect was indeed influenced by the distribution of faults in the evaluation benchmark, as described in the corresponding article~\cite{deepfd}. For artificial faults, \dfd, \NL, and \UM achieve equal RC performance, with \NL and \UM having higher PR and $F_3$ score. Overall, based on all the measured metrics, \dfd has the highest RC values, while \DD's RC measurements are noticeably lower than for other tools. \NL and \UM show similar performance according to the RC metric, while PR is slightly higher for \NL (0.23 vs 0.20), which achieved the highest values across all the considered tools. 

Despite the inferior performance of \DD in our experiments, the authors of \DD achieved higher performance for their tool than \UM in their evaluation. They used 2 separate sets of faulty programs. According to the results, \DD could correctly identify 87\% of the buggy models from one benchmark and 68\% from another, while \UM was only able to identify 49\% and 35\% of faulty models from these benchmarks, respectively. \UM's authors, in their turn, did not perform any empirical comparison with existing FL tools, and instead carried out a human study to measure how useful and usable is their tool for developers that aim to find and fix bugs in ML programs~\cite{schoop2021umlaut}. The authors of \NL also did not perform any comparison, but they have evaluated the performance of their tool on a set of 34 real-world faulty models gathered from SO posts and Github~\cite{nikanjam2021automatic}. Their evaluation showed that \NL was able to correctly detect 71\% of all the faults found in these issues. The authors of \dfd have performed their evaluation on a benchmark consisting of 58 real-world faults~\cite{deepfd}, that were later analysed by the authors of the benchmark~\cite{emp_repair} that we included in the real fault section for our study. Their evaluation showed that \dfd  can correctly localise 42\% of the ground truth faults observed in their benchmark, while \UM could only detect 23\%. 

It is worth mentioning that, unlike other tools, \dfd does not provide layer index suggestions. Thus, it is not possible to understand whether a successfully detected fault of `ACH' type actually points to the correct layer. This is the case for 2 issues out of 22 and if we exclude these issues from the calculation of average RC, the result for \dfd drops from 0.31 to 0.21, which makes it comparable with \NL and \UM. If we assume that \dfd correctly locates this fault with the probability of 50\% (the suggested layer is either correct or not), the mean RC value will be equal to 0.26. Also, for some of the fault types, other tools but \dfd provide specific suggestions on which activation function (DD, UM) or weights initialisation (NL) to adopt or whether to increase or decrease the learning rate (UM).

\begin{table*}[htb]
  \centering
  \caption{Number of Ground Truth (GT) faults (\#F); Recall (RC), Precision (PR) and $F_3$ measure for each FL tool. Avg. shows the average within artificial or real faults. T.A. shows the total average across faults.} 
  \label{tab:main_results_fl_2}
   \scalebox{0.8}{
  \begin{tabular}{l|c|ccc|ccc|ccc|ccc}
    \toprule
   Id & GT &\multicolumn{3}{c|}{DFD}&\multicolumn{3}{c|}{DD}&\multicolumn{3}{c|}{NL}&\multicolumn{3}{c}{UM}\\
     & \#F & RC  & PR & $F_3$ & RC  & PR & $F_3$ & RC & PR & $F_3$ & RC & PR & $F_3$ \\
    \midrule    
M1 & 1 &0  & 0&0& 0  & 0&0& 1  & 1&1& 0   & 0&0\\
M2 & 1 &0  & 0&0& 1  & 1&1& 0  & 0&0& 1   & 0.5&0.91\\
M3 & 1 &1  & 0.33&0.83& 0  & 0&0& 0  & 0&0& 0   & 0&0\\
C1 & 1 &1  & 0.25&0.77& 0  & 0&0& 0  & 0&0& 0   & 0&0\\
C2 & 1 &0  & 0&0& 0  & 0&0& 0  & 0&0& 0   & 0&0\\
C3 & 1 &0  & 0&0& 0  & 0&0& 1  & 1&1& 0   & 0&0\\
R1 & 1 &0  & 0&0& 0  & 0&0& 0  & 0&0& 0   & 0&0\\
R2 & 1 &0  & 0&0& 1  & 1&1& 0  & 0&0& 1   & 1&1\\
R3 & 1 &0  & 0&0& 0  & 0&0& 0  & 0&0& 1   & 1&1\\
R4 & 1 &1  & 0.50&0.91& 0  & 0&0& 1  & 1&1& 0   & 0&0\\
R5 & 1 &1  & 0.33&0.83& 0  & 0&0& 0  & 0&0& 0   & 0&0\\
R6 & 1 &0  & 0&0& 0  & 0&0& 1  & 1&1& 0   & 0&0\\
R7 & 1 &0  & 0&0& 1  & 1&1& 0  & 0&0& 1   & 1&1\\
    \midrule
\textbf{Avg.} & \textbf{1} &\textbf{0.31}  & \textbf{0.11} & \textbf{0.26} & \textbf{0.23}  & \textbf{0.23} & \textbf{0.23} & \textbf{0.31}  & \textbf{0.31} & \textbf{0.31} & \textbf{0.31}  & \textbf{0.27} & \textbf{0.30}\\
    \midrule
D1 & 1 &1  & 1&1& 0  & 0&0& 0  & 0&0& 0   & 0&0\\
D2 & 3 &0  & 0&0& 0  & 0&0& 0  & 0&0& 0   & 0&0\\
D3 & 5 &0.2  & 0.50&0.21& 0  & 0&0& 0.4  & 0.67&0.42& 0   & 0&0\\
D4 & 3 &0  & 0&0& 0.33  & 1&0.35& 0.33  & 0.5&0.34& 0.67   & 1&0.69\\
D5 & 2 &0  & 0&0& 0  & 0&0& 0  & 0&0& 0   & 0&0\\
D6 & 4 &0.5  & 0.67&0.51& 0  & 0&0& 0  & 0&0& 0   & 0&0\\
D7 & 1 &0  & 0&0& 0  & 0&0& 0  & 0&0& 0   & 0&0\\
D8 & 2 &1  & 0.50&0.91& 0  & 0&0& 0  & 0&0& 0   & 0&0\\
D9 & 3 &0  & 0&0& 0  & 0&0& 0  & 0&0& 0   & 0&0\\
    \midrule
\textbf{Avg.} & \textbf{2.67} &\textbf{0.30}  & \textbf{0.30}& \textbf{0.29}& \textbf{0.04}  & \textbf{0.11} & \textbf{0.04}& \textbf{0.08}  & \textbf{0.13} & \textbf{0.08} & \textbf{0.07}  & \textbf{0.11} & \textbf{0.08}\\ 			
    \midrule
\textbf{T.A.} & \textbf{1.68} & \textbf{0.31}  & \textbf{0.19} & \textbf{0.27} & \textbf{0.15}  & \textbf{0.18} &\textbf{0.15} & \textbf{0.22}  & \textbf{0.23} & \textbf{0.22} & \textbf{0.21}  & \textbf{0.20} & \textbf{0.21}\\
    \bottomrule
  \end{tabular}
  }
\end{table*}

\begin{tcolorbox}[colback = box-white]
  \textbf{RQ1 (before neutrality analysis)}: Our evaluation shows that all FL tools show relatively low RC results before neutrality analysis,  as, for many issues, the tools are not able to successfully identify the faults affecting the model according to the available ground truth. On average, \dfd shows the best results and \DD the lowest. At the same time, \NL and \UM perform quite similarly. 

\end{tcolorbox}

\subsection{RQ1 (Effectiveness after neutrality analysis)}

We have subsequently investigated our hypothesis that relying on a single ground truth, represented by a single set of changes that improve the model performance, may not be sufficient.

\begin{table*}[htb]
    \centering
  \caption{Neutrality Analysis}
  \label{tab:neutrality_stats}
    \begin{tabular}{c|c|c|c|c}
    \toprule
    Id & \# node  & \# alternative GTs & Complexity & Improvement \\
    \midrule
M1 & 258 & 240 & 3.54 (21) & 0.000 \\
M2 & 291 & 291 & 2.54 (21) & 0.000 \\
M3 & 170 & 170 & 2.01 (21) & 0.000 \\
C1 & 61 & 36 & 1.86 (27) & 0.003 \\
C2 & 31 & 1 & 1.00 (27) & 0.007 \\
C3 & 19 & 10 & 2.00 (27) & 0.004 \\
R1 & 45 & 0 & - (12) & - \\
R2 & 57 & 55 & 1.98 (12) & 0.008 \\
R3 & 60 & 0 & - (12) & - \\
R4 & 19 & 19 & 1.68 (12) & 0.009 \\
R5 & 31 & 19 & 2.58 (12) & 0.008 \\
R6 & 23 & 20 & 1.90 (12) & 0.004 \\
R7 & 38 & 38 & 2.79 (12) & 0.008 \\
\midrule
\textbf{Avg.} &\textbf{84.85} & \textbf{69.15} & \textbf{2.17 (18.55)} & \textbf{0.00} \\
\midrule
D1 & 92 & 92 & 4.29 (15) & 0.000 \\
D2 & 14 & 0 & - (21) & - \\
D3 & 47 & 44 & 8.59 (13) & 0.003 \\
D4 & 61 & 13 & 9.54 (12) & 0.010 \\
D5 & 41 & 1 & 4.00 (19) & 0.001 \\
D6 & 37 & 7 & 5.29 (12) & 0.000 \\
D7 & 49 & 25 & 4.04 (9) & 0.065 \\
D8 & 73 & 73 & 8.51 (17) & 0.186 \\
D9 & 22 & 0 & - (19) & - \\
\midrule
\textbf{Avg.} &\textbf{48.44} & \textbf{28.33} & \textbf{6.32 (13.86)} & \textbf{0.04} \\

            \bottomrule
    \end{tabular}
\end{table*}

Table~\ref{tab:neutrality_stats} shows the number of nodes in the neutrality graph, all of which are neutral relative to their parent nodes, and the number of found alternative GTs, which achieve equal or better performance than the known GT. Column `Complexity' shows how much the alternative GTs differ from the known GT, measured by counting how many hyperparameters differ between them. For each row, we calculate the complexity averaged over all found alternative GTs. The number in brackets indicates the total number of hyperparameters for each fault. Column `Improvement' shows the extent of performance improvement over the known GT in terms of the evaluation metric (e.g., accuracy), measured as the average difference across all  alternative GTs found. For four faults (R1, R3, D2, D9), it was not possible to identify alternative patches within the available budget and therefore no results could  be calculated (marked with `-'). The number of nodes varies depending on the dataset/model, with relatively smaller models such as MNIST (M1, M2, and M3) producing a more expanded network than others. 

Through our neutrality analysis, we identify an average of 69 alternative GTs for artificial faults and 28 for real faults, which reveals the existence of alternative GTs and could impact the evaluation of fault localisation tools. Typically, the complexity of real faults (6.32 on average) is higher than artificial faults (2.17 on average). This may stem from the fact that artificial faults are simpler by definition: by construction only one  hyperparameter is mutated, compared to the GT, whereas real faults tend to be more complex. In terms of the performance improvement of the alternative GTs over the known GT, we observe that there are only marginal improvements, although the improvements are more pronounced for  real faults compared to artificial faults. This could be attributed to the fact that the answers obtained from StackOverflow are not always ideal and may sometimes suggest only a partial fix.

Based on the results of  neutrality analysis, we have recalculated the fault localisation results for all the tools evaluated using the RC, PR or $F_3$ score.  Table~\ref{tab:neutrality_results_diff} shows results for each tool and issue that can be observed when using all the alternative ground truths, in addition to the original one. For issues, where it was not possible to locate alternative ground truths, results are greyed out. For issues on which a tool improved its performance after neutrality analysis, we indicate the improved RC, PR and $F_3$ score in boldface. The values of PR that have decreased in the result of taking alternative ground truth into consideration are underlined. In this table we report the maximum RC observed across all ground truth variants and the average PR and $F_3$ calculated on these GTs. 

It can be seen that \dfd is the tool that benefited the most from the alternative selection of ground truths, as its RC results have improved for 9 out of 18 issues for which the alternative GT was available. \dfd is followed by  \NL, whose RC results improved for 6 issues. On the other hand, for \DD and \UM, the RC values have increased in only 2 cases. To simplify the comparison of the tools before and after neutrality analysis, in Table~\ref{tab:neutrality_results} we provide initial average RC, PR and $F_3$ scores, along with the new ones, for the two benchmark sections (AF denotes artificial faults, RF real faults) and overall (T.A.: Total Average). It can be noticed that the  comparative performance observed across tools on the original ground truth is consistent with the one,  after neutrality analysis, despite the significant increase of the performance metrics exhibited by  \dfd and \NL. This is confirmed by the Wilcoxon signed-rank test with p-value of 0.002 for the comparison between \dfd and \DD and p-value of 0.023 for the \dfd vs \UM. The difference between \dfd and \NL is not statistically significant with p-value of 0.066.

Overall, our research highlights the importance of considering the existence of multiple potential fault-inducing changes. Fault localisation results change significantly when we broaden the definition of ground truth to include alternative fault-fixing changes.

\begin{table*}[htb]
  \centering
  \caption{Recall (RC), Precision (PR) and $F_3$ measure for each FL tool after neutrality analysis. Avg. shows the average within artificial or real faults. T.A. shows the total average across faults. The values that increased or decreased in comparison with the initial results (before neutrality analysis) are boldfaced or underlined, respectively. The issues for which neutrality analysis was not able to find any alternative ground truth are greyed out.} 
  \label{tab:neutrality_results_diff}
   \scalebox{0.8}{
  \begin{tabular}{l|ccc|ccc|ccc|ccc}
    \toprule
   Id &\multicolumn{3}{c|}{DFD}&\multicolumn{3}{c|}{DD}&\multicolumn{3}{c|}{NL}&\multicolumn{3}{c}{UM}\\
     & RC  & PR & $F_3$ & RC  & PR & $F_3$ & RC & PR & $F_3$ & RC & PR & $F_3$ \\
    \midrule    
M1&\textbf{0.67}& \textbf{0.5}&\textbf{0.65}& \textbf{0.5}& \textbf{1}&\textbf{0.53}& 1& 1&1.00& \textbf{0.5}& \textbf{1}&\textbf{0.53}\\
M2&\textbf{0.67}& \textbf{0.67}&\textbf{0.67}& 1& 1&1.00& \textbf{0.5}& \textbf{1}&\textbf{0.53}& 1& 0.5&0.91\\
M3&1& \textbf{0.47}&\textbf{0.88}& 0& 0&0& 0& 0&0& 0& 0&0\\
C1&1& \textbf{0.39}&\textbf{0.85}& 0& 0&0& 0& 0&0& 0& 0&0\\
C2&0& 0&0& 0& 0&0& 0& 0&0& 0& 0&0\\
C3&\textbf{1}& \textbf{0.25}&\textbf{0.77}& 0& 0&0& 1& 1&1& 0& 0&0\\
\rowcolor{lightgray}
R1&0& 0&0& 0& 0&0& 0& 0&0& 0& 0&0\\
R2&\textbf{1}& \textbf{0.56}&\textbf{0.91}& 1& 1&1& \textbf{1}& \textbf{1}&\textbf{1}& 1& 1&1\\
\rowcolor{lightgray}
R3&0& 0&0& 0& 0&0& 0& 0&0& 1& 1&1\\
R4&1& \textbf{0.57}&\textbf{0.92}& 0& 0&0& 1& 1&1& 0& 0&0\\
R5&1& \textbf{0.5}&\textbf{0.89}& 0& 0&0& 0& 0&0& 0& 0&0\\
R6&\textbf{0.5}& \textbf{0.25}&\textbf{0.45}& 0& 0&0& 1& 1&1& 0& 0&0\\
R7&\textbf{1}& \textbf{0.5}&\textbf{0.89}& 1& 1&1& \textbf{1}& \textbf{1}&\textbf{1}& 1& 1&1\\
 \textbf{Avg.}& \textbf{0.68}& \textbf{0.36}& \textbf{0.61}& \textbf{0.27}& \textbf{0.28}& \textbf{0.27}& \textbf{0.50}& \textbf{0.54}& \textbf{0.50}& \textbf{0.35}& \textbf{0.35}& \textbf{0.34}\\
    \midrule
D1&1& 1&1& \textbf{0.5}& \textbf{1}&\textbf{0.53}& \textbf{0.5}& \textbf{1}&\textbf{0.53}& 0& 0&0\\
\rowcolor{lightgray}
D2&0& 0&0& 0& 0&0& 0& 0&0& 0& 0&0\\
D3&\textbf{1}& 0.5&\textbf{0.91}& 0& 0&0& \textbf{0.5}& \underline{0.33}&\textbf{0.48}& 0& 0&0\\
D4&0& 0&0& 0.33& 1&0.35& \textbf{1}& 0.5&\textbf{0.91}& \textbf{1}& \underline{0.5}&\textbf{0.91}\\
D5&0& 0&0& 0& 0&0& 0& 0&0& 0& 0&0\\
D6& \textbf{1} & \underline{0.5} &\textbf{0.89}& 0& 0&0& 0& 0&0& 0& 0&0\\
D7&\textbf{0.5}& \textbf{1}&\textbf{0.53}& 0& 0&0& 0& 0&0& 0& 0&0\\
D8&1& 0.5&0.91& 0& 0&0& 0& 0&0& 0& 0&0\\
\rowcolor{lightgray}
D9&0& 0&0& 0& 0&0& 0& 0&0& 0& 0&0\\
 \textbf{Avg.}& 0.50& 0.39& 0.47& 0.09& 0.22& 0.10& 0.22& 0.20& 0.21& 0.11& 0.06& 0.10\\
 \midrule
 \textbf{T.A.}& \textbf{0.61}& \textbf{0.37}& \textbf{0.55}& \textbf{0.20}& \textbf{0.26}& \textbf{0.20}& \textbf{0.39}& \textbf{0.40}& \textbf{0.38}& \textbf{0.25}& \textbf{0.23}& \textbf{0.24}\\
    \bottomrule
  \end{tabular}
  }
\end{table*}

\begin{table*}[htb]
  \centering
  \caption{Overall comparison of Recall (RC), Precision (PR) and $F_3$ measure for each FL tool before/after neutrality analysis. Avg. shows the average within artificial (AF) or real (RF) faults. T.A. shows the total average across faults.} 
  \label{tab:neutrality_results}
   \scalebox{0.8}{
  \begin{tabular}{l|ccc|ccc|ccc|ccc}
    \toprule
   Id &\multicolumn{3}{c|}{DFD}&\multicolumn{3}{c|}{DD}&\multicolumn{3}{c|}{NL}&\multicolumn{3}{c}{UM}\\
     & RC  & PR & $F_3$ & RC  & PR & $F_3$ & RC & PR & $F_3$ & RC & PR & $F_3$ \\
    \midrule
 \multicolumn{13}{c}{Before neutrality analysis}\\
 \midrule
 \textbf{AF Avg.}& 0.31& 0.11& 0.26& 0.23& 0.23& 0.23& 0.31& 0.31& 0.31& 0.31& 0.27&0.3\\
 \textbf{RF Avg.}& 0.3& 0.3& 0.29& 0.04& 0.11& 0.04& 0.08& 0.13& 0.08& 0.07& 0.11&0.08\\
 \textbf{T.A.} & 0.31& 0.19& 0.27& 0.15& 0.18& 0.15& 0.22& 0.23& 0.22& 0.21& 0.2&0.21\\
 \midrule
 \multicolumn{13}{c}{After neutrality analysis}\\
 \midrule
\textbf{AF Avg.}&0.68& 0.36& 0.61& 0.27& 0.28& 0.27& 0.50& 0.54& 0.50& 0.35& 0.35& 0.34\\
\textbf{RF Avg.}&0.50& 0.48& 0.48& 0.09& 0.22& 0.10& 0.22& 0.20& 0.21& 0.11& 0.06& 0.10\\ 			
\textbf{T.A.} & 0.61& 0.41& 0.55& 0.20& 0.26&0.20& 0.39& 0.40& 0.38& 0.25& 0.23& 0.24\\
    \bottomrule
  \end{tabular}
  }
\end{table*}

\begin{tcolorbox}[colback = box-white]
  \textbf{RQ1 (after neutrality analysis)}: Our evaluation after neutrality analysis shows that the performance of all FL tools increase if alternative ground truths are considered, with \dfd and \NL exhibiting the largest improvements. Still, the  relatively low RC even after neutrality analysis indicates that DL fault localization is still and open problem, requiring  future research. They also indicate the fundamental importance of considering alternative ground truths in the evaluation of DL fault localization tools.
\end{tcolorbox}

\subsection{RQ2 (Stability)}

The authors of \dfd account for the instability of the training process and perform 20 retrainings when collecting input features both during the classifier training stage and during fault identification. This way, the output of the tool is calculated from 20 feature sets for each model under test.

\NL does not require any training and is based on static rules that are stable by design. We performed 20 runs of all other tools to investigate their stability. We found out that outputs are stable across the experiment repetitions for all considered tools.

\begin{tcolorbox}[colback = box-white]
  \textbf{RQ2}: Existing fault localisation tools provide stable results that do not change from execution to execution.
\end{tcolorbox}

\subsection{RQ3 (Efficiency)}

In this RQ, we investigate how demanding the evaluated approaches are in terms of execution time. Here we measure only the time required to run an FL tool on a subject, without taking into account the time and effort needed to prepare the subject for the tool application. All of the tools require some manual work to be done: for \dfd, a user has to create serialised versions of the training dataset and model configuration according to a specific format; for \DD and \UM, a user has to insert  a tool-specific callback to the code and provide it with a list of arguments; for \NL, there is a number of manual changes to the source code to make the tool runnable.

Table~\ref{tab:main_results_fl_3} shows execution time measured in seconds on a single run of \dfd and \NL, and the average of 20 runs for the remaining tools. Row `T.A.' shows the average time spent by each tool on fault localisation over the whole benchmark. To allow fair comparison, row `Avg.' shows the average execution time over the faults where all tools are applicable. Not surprisingly, \dfd takes considerably longer to run than the other tools since it trains 20 instances for each issue, while the other tools perform one (\DD, \UM) or no retraining (\NL). In addition, \DD often terminates the training when a faulty behaviour is observed, which makes its average execution time the shortest on the issues we used. As \NL does not require to train a model to perform fault localisation, its average execution time is also quite low. It can be noted that for some faults that are very fast to train (e.g. C2, D1, D3, 6), a full training performed by \UM takes less time than the static checks of \NL. On average, \DD is the fastest to run, followed by \NL and \UM, and finally, \dfd. Despite the differences, the execution time of all considered tools  is compatible with real-world use. In Figure~\ref{fig:f3time}, we show the average execution time of each tool combined with the average performance measured using the $F_3$ score. A longer execution time is paid off in terms of higher effectiveness in the case of \dfd, while this is not the case of \DD and \UM, which are outperformed by the extremely efficient \NL.

\begin{table*}[htb]
  \centering
  \caption{Execution time (in seconds)} 
  \label{tab:main_results_fl_3}
   \scalebox{1}{
  \begin{tabular}{l|c|c|c|c}
    \toprule
    ID & DFD & DD & NL & UM \\
    \midrule
    M1 & 605.30 & 6.65 & 7.63 & 37.62  \\
    M2 & 485.34 & 6.84 & 9.95 & 40.17  \\
    C1 & 316.10 & 7.34 & 10.02 & 163.08  \\
    C2 & 338.45 & 7.15 & 9.77 & 4.77  \\
    C3 & 321.42 & 7.03 & 10.02 & 135.75  \\
    R1 & 124.50 & 4.75 & 9.44 & 6.25  \\
    R2 & 115.12 & 4.05 & 9.59 & 5.89  \\
    R4 & 125.76 & 3.90 & 9.59 & 6.16  \\
    R5 & 126.13 & 3.58 & 7.7 & 5.10  \\
    R6 & 133.23 & 4.07 & 9.19 & 6.02  \\
    R7 & 158.34 & 3.95 & 8.99 & 6.07  \\
    D1 & 54.50 & 3.30 & 9.85 & 2.07  \\
    D2 & 451.67 & 20.13 & 9.95 & 18.98  \\
    D3 & 32.80 & 1.58 & 9.50 & 1.33  \\
    D4 & 797.46 & 11.66 & 6.87 & 324.57  \\
    D5 & 562.46 & 11.54 & 7.50 & 27.43  \\
    D6 & 19.6 & 1.32 & 6.88 & 0.39  \\
    D8 & 109.40 & 2.36 & 10.16 & 4.38  \\
    \midrule
    \textbf{Avg.} & \textbf{270.98} & \textbf{6.18} & \textbf{9.03} & \textbf{44.22}  \\
    \midrule
    M3 & 798.23 & 6.86 & N/A & 38.66  \\
    R3 & 116.34 & 4.05 & N/A & 6.00  \\
    D7 & 53.53 & 166.12 & N/A & 1.89  \\
    D9 & N/A & N/A & 9.35 & 57.07  \\
    \midrule
\textbf{T.A.} & \textbf{278.37} & \textbf{13.73} & \textbf{9.05} & \textbf{40.89} \\
    \bottomrule
  \end{tabular}
  }
\end{table*}

\begin{figure}[htb]
    \centering
    \includegraphics[width=1\linewidth]{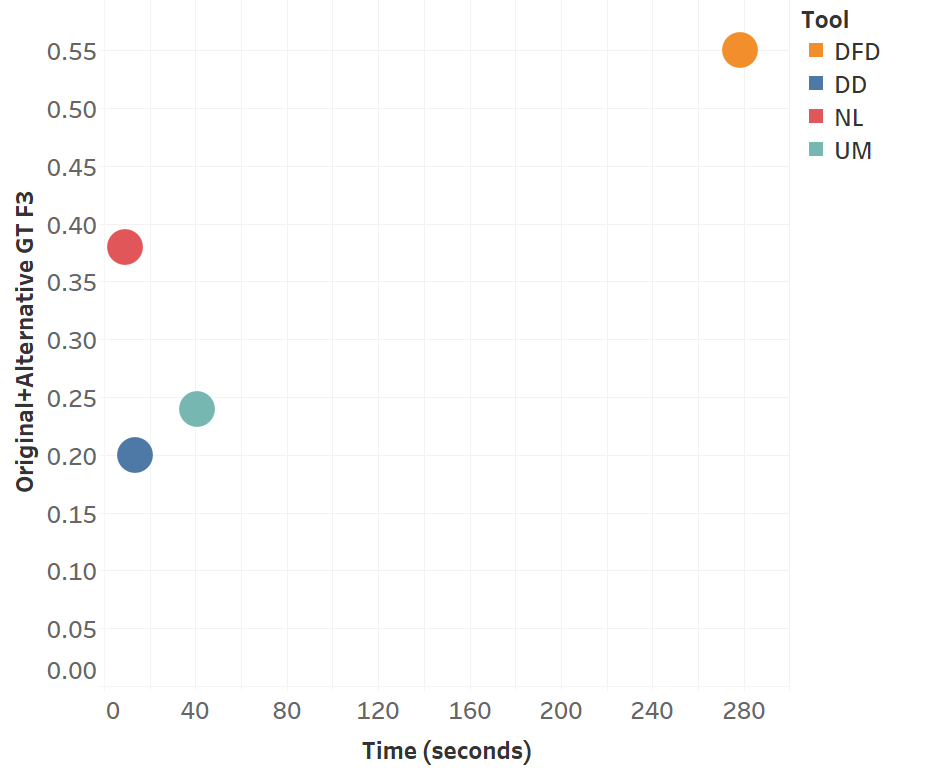}
    \caption{Average execution time and average performance ($F_3$ score) for each tool}
    \label{fig:f3time}
\end{figure}

\begin{tcolorbox}[colback = box-white]
  \textbf{RQ3}: The tools considered in our empirical study operate on the basis of different strategies and require different numbers of retrainings of the same model to attempt fault localisation. \dfd is the slowest, as it trains 20 instances of the model under test, but it is also the most effective. \NL 
 performs fault localisation without training a model, making it often faster and on average more effective than \UM (as well as \DD). No tool requires an execution that is prohibitively expensive for practical use.
\end{tcolorbox}

\subsection{The Outputs of Fault Localisation Tools}

The analysed fault localisation tools output messages in  natural language that explain where the fault is present in the neural network. The level of understandability of these messages, as well as the degree of detail they provide about the location of the fault and its possible fixes, is an important indicator of the applicability of these tools in practice. We have analysed each output message produced by each tool and provide an overview of our findings in this subsection.

The output of \dfd consists of a vector of parameters from the following list [\textit{`optimizer', `lr', `loss', `epoch', `act'}]. In our experiments, the size of the output vector varied between 1 and 4. While the provided vector clearly indicates which hyperparameters might contain faulty values, it provides no indication on how the values should be changed, e.g., whether the learning rate should be increased or decreased. Moreover, for the activation hyperparameter, the layer number for which the activation function should be changed is not indicated. 

The outputs of \DD often refer to problems observed for internal parameters during the training process. Some of these outputs, such as "\textit{Batch 0 layer 6: Numerical Error in delta Weights, terminating training}", do not provide any guidance on what should be changed in the model architecture, training data or hyperparameters to fix the fault. In contrast, some other outputs such as "\textit{Batch 0 layer 9:  Out of Range Problem, terminating training. Change the activation function to softmax}" or "\textit{Batch 0 layer 0: Vanishing Gradient Problem in delta Weights, terminating training. Add/delete layer or change sigmoid activation function}" are more instructive and provide layer numbers along with the required changes.

\UM provides an output that lists critical issues as well as warnings, e.g., "\textit{<Critical: Missing Softmax layer before loss>, <Warning: Last model layer has nonlinear activation>}" and "\textit{<Critical: Missing Softmax layer before loss>, <Critical: Missing activation functions>, <Warning: Last model layer has nonlinear activation>}". It should be noted that in our experiments \UM reported the critical issue of "\textit{Missing Softmax layer before loss}" for all the analysed faults, including the cases when the softmax layer is already present in the model architecture. Messages indicate the layer number ("before loss" or "last model layer") in some cases, while in others this information is missing ("\textit{<Critical: Missing activation functions>}"). Similarly to \DD, some of the warnings produced by \UM do not contain actionable fix suggestions, for example, "Possible over-fitting" or "Check validation accuracy".

The output messages of \NL report faults either in specific layers ("\textit{Layer 4 ==> The initialization of weights should not be constant to break the symmetry between neurons}") or in the learning process ("\textit{Learner ==> The loss should be correctly defined and connected to the layer in accordance with its input conditions (i.e., shape and type)-post\_activation}"). The messages provide information on which component is faulty, along with an explanation of why it is faulty. 

Overall, except for some cases in \DD, the outputs of FL tools provide clear messages indicating which types of hyperparameters are faulty. However, in cases when the hyperparameter can be applied to different layers of the model, the localisation to the specific layer is not always performed. When it comes to fix suggestions, while \dfd provides no information in this direction, the remaining tools have some output messages that come with suggested repairs.

\section{Threats to Validity}
\label{sec:threats}

\subsection{Construct}
Threats to construct validity are due to the measurement of the
effectiveness of the fault localisation tools and the interpretation of the tools' output. We use a simple count of the matches between fault localisation results and the ground truth, along with the RC, PR and $F_{\beta}$ metrics that are quite standard in information retrieval.

\subsection{Internal}
Threats to the internal validity of the study lie in the selection of evaluated approaches. To the best of our knowledge, we considered all state-of-the-art techniques and adopted their publicly available implementations.

\subsection{External}
To address the threats to external validity, we carefully selected faults of both artificial and real nature, covering a set of diverse subjects for the evaluation of FL techniques. Nevertheless, replicating our study on additional subjects would be useful to corroborate our findings.

\section{Related Work}
\label{sec:related_work}

While to the best of our knowledge ours is the first empirical study that performs a third party assessment of existing DL fault localization tools, there is a previous empirical work~\cite{emp_repair} aimed at comparing different DL repair approaches. In the following, we first discuss such empirical work, followed by a summary presentation of existing repair approaches: although they do not address the DL fault localization problem, they are relevant to such task.

The DNN model architecture repair problem, as defined in the recent study by Kim et al.~\cite{emp_repair}, lies in improving the performance of a faulty deep neural network (DNN) model by finding an alternative configuration of its architecture and hyperparemeters. The new configuration should lead to a statistically significant enhancement in model performance, such as accuracy or mean squared error, when measured on a test dataset. In particular, the authors consider a number of  categories and subcategories from a DL fault taxonomy~\cite{taxonomy},  covering the following issues: faults affecting the structure and properties, faults affecting the DNN layer properties and activation
functions, faults due to missing/redundant/wrong layers, and
faults associated with the choice of optimiser, loss function and
hyperparameters (e.g., learning rate, number of epochs) as model architecture faults. Examples of such faults include the selection of an inappropriate loss function for the task at hand or training a model for an insufficient number of epochs. 

Existing advances in Hyperparameter Optimization (HPO) can be considered as a way to address the problem of repair as they can be applied to search optimal configurations for different aspects of model architecture such as activation functions, number of neurons and layers, hyperparameters affecting the training process, etc. At the moment, there is no automated \textit{source-level} repair tool that improves performance of a model by means of patching and modifying the sources of the model's architecture. However, there exists a tool called AutoTrainer~\cite{autotrainer} which is designed to detect and repair training problems such as dying ReLU or exploding gradients, by continuing the training with patched architecture or hyperparameters.

Kim et al.~\cite{emp_repair} compared AutoTrainer with HEBO~\cite{Cowen-Rivers2022lm} and BOHB~\cite{bohb}, state-of-the-art HPO techniques based on Bayesian Optimisation (BO), while using random search as a baseline. The comparison was performed on a carefully compiled set of artificial and real-world faulty models. Their results demonstrate that the evaluated techniques can potentially improve the performance of models affected by architecture faults. However, their findings indicate that there is still considerable room for improvement as random baseline performs quite well when compared with other techniques.
 
On the other hand, there exist a number of post-training \textit{model-level} repair approaches that focus on modifying the weights of an already trained model in order to eliminate observed misbehavours. 
Arachne~\cite{arachne} and Care~\cite{sun2022causality} both focus on identifying the neurons that contribute the most to the detected misbehaviours on certain test inputs, and calibrate the weights associated with these neurons, while trying not to corrupt correct predictions. GenMuNN~\cite{wu2022genmunn}, however, directly locates the weights that play the biggest role in predictions, and uses a genetic algorithm to evolve the model by applying slight mutations to such weights. I-Repair~\cite{henriksen2022repairing} also locates and changes the weights that take part in forming a misbehaving output for a certain group of inputs, while maintaining the same behaviour on correctly classified inputs. PRDNN~\cite{sotoudeh2021provable} similarly aims at producing the smallest achievable single-layer repair. NNrepair~\cite{usman2021nn} uses constraint solving to produce slight modifications to suspicious weights revealed by fault localisation. Apricot~\cite{apricot} adjusts the weights of a misbehaving model using the guidance from the weights of a complementary correctly-performing model trained on reduced dataset that contains the problematic inputs.

While hyperparameter optimization tools provide source level information about the performed fixes, which means they also offer some fault localization capability, post-training repair tools are completely opaque and their fixes have no interpretation in terms of architectural model elements affected by a fault. In our empirical study, we restricted the selection of tools to those that explicitly address the DL fault localization problem.

\section{Conclusion}
\label{sec:conclusion}

We evaluated four state-of-the-art techniques in DL fault localisation on a  meticulously tailored set of real and artificial faulty models to assess the advances in the area. Our findings show that all of the evaluated approaches are able to locate a certain percentage of faults. However, all are quite far from the best possible results when considering the available ground truth. \dfd exhibited the highest effectiveness, followed by \NL and \UM. \DD exhibited relatively poor performance. On the positive side, all proposed techniques are stable across multiple runs and do not require excessive execution time. However, our experimentation suggests that when re-computing the results after including multiple alternative to ground truth patches (obtained by neutrality analysis), the FL accuracy of tools increases  in all cases, sometimes quite substantially.

According to our findings, future work in the area of DL fault  localisation should focus on improving the fault identification capabilities of the proposed techniques and broadening the variety of considered fault types. Moreover, any empirical evaluation of DL fault localisation tools should include some form of neutrality analysis, to expand the available ground truth to other possible, equivalent fixes.

\section{Conflict of Interest and Data Availability} \label{sec:data}
All authors confirm that they are not affiliated with or involved in any organisation or entity that has financial or non-financial interests in the subject matter or materials discussed in this manuscript.

The experimental data, and the evaluation results that support the findings of this study are available in Zenodo platform under the following identifier: 10.5281/zenodo.10387016.

\section*{Acknowledgements}
Jinhan Kim and Shin Yoo have been supported by the Engineering Research Center
Program through the National Research Foundation of Korea (NRF) funded by the
Korean Government (MSIT) (NRF-2018R1A5A1059921), NRF Grant (NRF-2020R1A2C1013629),
Institute for Information \& communications Technology Promotion grant funded by
the Korean government (MSIT) (No.2021-0-01001), and Samsung Electronics (Grant
No. IO201210-07969-01). This work was partially supported by the H2020 project
PRECRIME, funded under the ERC Advanced Grant 2017 Program (ERC Grant Agreement
n. 787703).

\bibliographystyle{IEEEtran}
\balance

\bibliography{biblio2}

\begin{thebibliography}{10}
\providecommand{\url}[1]{#1}
\csname url@samestyle\endcsname
\providecommand{\newblock}{\relax}
\providecommand{\bibinfo}[2]{#2}
\providecommand{\BIBentrySTDinterwordspacing}{\spaceskip=0pt\relax}
\providecommand{\BIBentryALTinterwordstretchfactor}{4}
\providecommand{\BIBentryALTinterwordspacing}{\spaceskip=\fontdimen2\font plus
\BIBentryALTinterwordstretchfactor\fontdimen3\font minus
  \fontdimen4\font\relax}
\providecommand{\BIBforeignlanguage}[2]{{%
\expandafter\ifx\csname l@#1\endcsname\relax
\typeout{** WARNING: IEEEtran.bst: No hyphenation pattern has been}%
\typeout{** loaded for the language `#1'. Using the pattern for}%
\typeout{** the default language instead.}%
\else
\language=\csname l@#1\endcsname
\fi
#2}}
\providecommand{\BIBdecl}{\relax}
\BIBdecl

\bibitem{deeplocalize}
\BIBentryALTinterwordspacing
M.~Wardat, W.~Le, and H.~Rajan, ``Deeplocalize: Fault localization for deep
  neural networks,'' in \emph{2021 IEEE/ACM 43rd International Conference on
  Software Engineering (ICSE)}.\hskip 1em plus 0.5em minus 0.4em\relax Los
  Alamitos, CA, USA: IEEE Computer Society, may 2021, pp. 251--262. [Online].
  Available:
  \url{https://doi.ieeecomputersociety.org/10.1109/ICSE43902.2021.00034}
\BIBentrySTDinterwordspacing

\bibitem{wardat2022deepdiagnosis}
M.~Wardat, B.~D. Cruz, W.~Le, and H.~Rajan, ``{DeepDiagnosis}: automatically
  diagnosing faults and recommending actionable fixes in deep learning
  programs,'' in \emph{Proceedings of the 44th International Conference on
  Software Engineering}, 2022, pp. 561--572.

\bibitem{deepfd}
\BIBentryALTinterwordspacing
J.~Cao, M.~Li, X.~Chen, M.~Wen, Y.~Tian, B.~Wu, and S.-C. Cheung, ``Deepfd:
  Automated fault diagnosis and localization for deep learning programs,'' in
  \emph{Proceedings of the 44th International Conference on Software
  Engineering}, ser. ICSE '22.\hskip 1em plus 0.5em minus 0.4em\relax New York,
  NY, USA: Association for Computing Machinery, 2022, p. 573–585. [Online].
  Available: \url{https://doi.org/10.1145/3510003.3510099}
\BIBentrySTDinterwordspacing

\bibitem{nikanjam2021automatic}
A.~Nikanjam, H.~B. Braiek, M.~M. Morovati, and F.~Khomh, ``Automatic fault
  detection for deep learning programs using graph transformations,'' \emph{ACM
  Transactions on Software Engineering and Methodology (TOSEM)}, vol.~31,
  no.~1, pp. 1--27, 2021.

\bibitem{schoop2021umlaut}
E.~Schoop, F.~Huang, and B.~Hartmann, ``Umlaut: Debugging deep learning
  programs using program structure and model behavior,'' in \emph{Proceedings
  of the 2021 CHI Conference on Human Factors in Computing Systems}, 2021, pp.
  1--16.

\bibitem{autotrainer}
X.~Zhang, J.~Zhai, S.~Ma, and C.~Shen, ``Autotrainer: An automatic dnn training
  problem detection and repair system,'' in \emph{2021 IEEE/ACM 43rd
  International Conference on Software Engineering (ICSE)}, 2021, pp. 359--371.

\bibitem{bakerdetect}
W.~Baker, M.~O’Connor, S.~R. Shahamiri, and V.~Terragni, ``Detect, fix, and
  verify tensorflow api misuses,'' in \emph{International Conference on
  Software Analysis, Evolution and Reengineering}, 2022, pp. 1--5.

\bibitem{keras}
``Keras,'' Available at \url{https://keras.io}.

\bibitem{Zhang:2018}
\BIBentryALTinterwordspacing
Y.~Zhang, Y.~Chen, S.-C. Cheung, Y.~Xiong, and L.~Zhang, ``An empirical study
  on tensorflow program bugs,'' in \emph{Proceedings of the 27th ACM SIGSOFT
  International Symposium on Software Testing and Analysis}, ser. ISSTA
  2018.\hskip 1em plus 0.5em minus 0.4em\relax New York, NY, USA: ACM, 2018,
  pp. 129--140. [Online]. Available:
  \url{http://doi.acm.org/10.1145/3213846.3213866}
\BIBentrySTDinterwordspacing

\bibitem{knearestn}
\BIBentryALTinterwordspacing
N.~S. Altman, ``An introduction to kernel and nearest-neighbor nonparametric
  regression,'' \emph{The American Statistician}, vol.~46, no.~3, pp. 175--185,
  1992. [Online]. Available: \url{http://www.jstor.org/stable/2685209}
\BIBentrySTDinterwordspacing

\bibitem{breiman2017classification}
L.~Breiman, \emph{Classification and regression trees}.\hskip 1em plus 0.5em
  minus 0.4em\relax Routledge, 2017.

\bibitem{ho1995random}
T.~K. Ho, ``Random decision forests,'' in \emph{Proceedings of 3rd
  international conference on document analysis and recognition}, vol.~1.\hskip
  1em plus 0.5em minus 0.4em\relax IEEE, 1995, pp. 278--282.

\bibitem{deepcrime}
\BIBentryALTinterwordspacing
N.~Humbatova, G.~Jahangirova, and P.~Tonella, ``Deepcrime: Mutation testing of
  deep learning systems based on real faults,'' in \emph{Proceedings of the
  30th ACM SIGSOFT International Symposium on Software Testing and Analysis},
  ser. ISSTA 2021.\hskip 1em plus 0.5em minus 0.4em\relax New York, NY, USA:
  Association for Computing Machinery, 2021, p. 67–78. [Online]. Available:
  \url{https://doi.org/10.1145/3460319.3464825}
\BIBentrySTDinterwordspacing

\bibitem{taxonomy}
N.~Humbatova, G.~Jahangirova, G.~Bavota, V.~Riccio, A.~Stocco, and P.~Tonella,
  ``Taxonomy of real faults in deep learning systems,'' in \emph{Proceedings of
  the 41st International Conference on Software Engineering, ICSE}, 2020.

\bibitem{neutrality}
\BIBentryALTinterwordspacing
J.~Renzullo, W.~Weimer, M.~Moses, and S.~Forrest, ``Neutrality and epistasis in
  program space,'' in \emph{Proceedings of the 4th International Workshop on
  Genetic Improvement Workshop}, ser. GI '18.\hskip 1em plus 0.5em minus
  0.4em\relax New York, NY, USA: Association for Computing Machinery, 2018, p.
  1–8. [Online]. Available: \url{https://doi.org/10.1145/3194810.3194812}
\BIBentrySTDinterwordspacing

\bibitem{emp_repair}
J.~Kim, N.~Humbatova, G.~Jahangirova, P.~Tonella, and S.~Yoo, ``Repairing dnn
  architecture: Are we there yet?'' in \emph{2023 IEEE Conference on Software
  Testing, Verification and Validation (ICST)}, 2023.

\bibitem{deepfd_replication}
J.~Cao, M.~Li, X.~Chen, M.~Wen, Y.~Tian, B.~Wu, and S.-C. Cheung, ``Replication
  package of deepfd,'' \url{https://github.com/ArabelaTso/DeepFD}, 2021.

\bibitem{umlaut_replication}
E.~Schoop, F.~Huang, and B.~Hartmann, ``Replication package of umlaut,''
  Available at \url{https://github.com/BerkeleyHCI/umlaut}, 2021.

\bibitem{neuralint_replication}
A.~Nikanjam, H.~B. Braiek, M.~M. Morovati, and F.~Khomh, ``Replication package
  of {Neuralint},'' Available at \url{https://github.com/neuralint/neuralint}.

\bibitem{deepdiagnosis_replication}
M.~Wardat, B.~D. Cruz, W.~Le, and H.~Rajan, ``Replication package of
  {DeepDiagnosis},'' Available at
  \url{https://github.com/deepdiagnosis/icse2022}, 2021.

\bibitem{JahangirovaICST20}
\BIBentryALTinterwordspacing
G.~Jahangirova and P.~Tonella, ``An empirical evaluation of mutation operators
  for deep learning systems,'' in \emph{IEEE International Conference on
  Software Testing, Verification and Validation}, ser. ICST'20.\hskip 1em plus
  0.5em minus 0.4em\relax IEEE, 2020, p. 12 pages. [Online]. Available:
  \url{https://doi.org/10.1109/ICST46399.2020.00018}
\BIBentrySTDinterwordspacing

\bibitem{fl_comparison_replication}
``{Empirical} {Comparison} of {Fault} {Localisation} {Techniques} for {DNNs}
  (replication package),'' Available at
  \url{https://zenodo.org/doi/10.5281/zenodo.10387015}, 2023.

\bibitem{Cowen-Rivers2022lm}
A.~I. Cowen-Rivers, W.~Lyu, R.~Tutunov, Z.~Wang, A.~Grosnit, R.~R. Griffiths,
  A.~M. Maraval, H.~Jianye, J.~Wang, J.~Peters \emph{et~al.}, ``Hebo: Pushing
  the limits of sample-efficient hyper-parameter optimisation,'' \emph{Journal
  of Artificial Intelligence Research}, vol.~74, pp. 1269--1349, 2022.

\bibitem{bohb}
S.~Falkner, A.~Klein, and F.~Hutter, ``Bohb: Robust and efficient
  hyperparameter optimization at scale,'' in \emph{International Conference on
  Machine Learning}.\hskip 1em plus 0.5em minus 0.4em\relax PMLR, 2018, pp.
  1437--1446.

\bibitem{arachne}
\BIBentryALTinterwordspacing
J.~Sohn, S.~Kang, and S.~Yoo, ``Arachne: Search based repair of deep neural
  networks,'' \emph{ACM Trans. Softw. Eng. Methodol.}, sep 2022. [Online].
  Available: \url{https://doi.org/10.1145/3563210}
\BIBentrySTDinterwordspacing

\bibitem{sun2022causality}
B.~Sun, J.~Sun, L.~H. Pham, and J.~Shi, ``Causality-based neural network
  repair,'' in \emph{Proceedings of the 44th International Conference on
  Software Engineering}, 2022, pp. 338--349.

\bibitem{wu2022genmunn}
H.~Wu, Z.~Li, Z.~Cui, and J.~Liu, ``Genmunn: A mutation-based approach to
  repair deep neural network models,'' \emph{International Journal of Modeling,
  Simulation, and Scientific Computing}, p. 2341008, 2022.

\bibitem{henriksen2022repairing}
P.~Henriksen, F.~Leofante, and A.~Lomuscio, ``Repairing misclassifications in
  neural networks using limited data,'' in \emph{Proceedings of the 37th
  ACM/SIGAPP Symposium on Applied Computing}, 2022, pp. 1031--1038.

\bibitem{sotoudeh2021provable}
M.~Sotoudeh and A.~V. Thakur, ``Provable repair of deep neural networks,'' in
  \emph{Proceedings of the 42nd ACM SIGPLAN International Conference on
  Programming Language Design and Implementation}, 2021, pp. 588--603.

\bibitem{usman2021nn}
M.~Usman, D.~Gopinath, Y.~Sun, Y.~Noller, and C.~S. P{\u{a}}s{\u{a}}reanu, ``Nn
  repair: Constraint-based repair of neural network classifiers,'' in
  \emph{International Conference on Computer Aided Verification}.\hskip 1em
  plus 0.5em minus 0.4em\relax Springer, 2021, pp. 3--25.

\bibitem{apricot}
H.~Zhang and W.~Chan, ``Apricot: A weight-adaptation approach to fixing deep
  learning models,'' in \emph{2019 34th IEEE/ACM International Conference on
  Automated Software Engineering (ASE)}, 2019, pp. 376--387.

\end{thebibliography}

\vspace{12pt}
\end{document}